# Layer-engineered quantum anomalous Hall effect in twisted rhombohedral graphene

Zhangyuan Chen[1,2*], Naitian Liu[1,2*], Jiannan Hua[1,2*], Hanxiao Xiang[1,2], Wenqiang Zhou[1,2], Jing Ding[1,2], Xinjie Fang[1,2], Linfeng Wu[1,2], Le Zhang[1,2], Qianmei Chen[3], Xuanyu Chen[3], Kenji Watanabe[4], Takashi Taniguchi[5], Na Xin[3], Wei Zhu[1,2†], Shuigang Xu[1,2†]

[1] *Key Laboratory for Quantum Materials of Zhejiang Province, Department of Physics, School of Science, Westlake University, Hangzhou 310030, China*

[2] *Institute of Natural Sciences, Westlake Institute for Advanced Study, Hangzhou 310024, China*

[3] *Department of Chemistry, Zhejiang University, Hangzhou 310058, China*

[4] *Research Center for Electronic and Optical Materials, National Institute for Materials Science, Tsukuba 305-0044, Japan*

[5] *Research Center for Materials Nanoarchitectonics, National Institute for Materials Science, Tsukuba 305-0044, Japan*

[*]These authors contributed equally to this work.

[†]Corresponding authors: zhuwei@westlake.edu.cn (W.Z.); xushuigang@westlake.edu.cn (S.X.)

**Abstract**

Realizing programmable topological states in quantum anomalous Hall (QAH) insulators requires the ability to design and dynamically tune their topological invariant, the Chern number $C$. Here, we report a designer QAH platform based on twisted rhombohedral graphene family, in which $C$ becomes a programmable and electrically tunable degree of freedom. By engineering the layer configuration in twisted monolayer-rhombohedral $N$-layer graphene, denoted as $(1+N)$ L, we realize QAH states with $C = N$ at moiré filling $\nu = 1$, where the layer number $N = 3, 4, 5$ directly sets the Chern number. Beyond such static layer programming, we demonstrate *in-situ* electrical control. In a twisted monolayer-trilayer device, the sign of $C$ (chirality) can be switched by electrostatic doping or displacement field. Most strikingly, in twisted Bernal bilayer-rhombohedral tetralayer graphene denoted as $(2+4)$ L, we drive a displacement-field-induced topological phase transition between two distinct QAH states with $C = 3$ and $C = 4$ in a single device. Our work establishes a layer-engineered and electrically tunable platform that transitions topological quantum matter from discovery to design, opening the way toward on-demand engineering of correlated topological states and reconfigurable topological electronics.

The quantum anomalous Hall (QAH) effect, which exhibits a quantized Hall resistance of $R_{xy} = h/Ce^2$ and vanishing longitudinal resistance in the absence of an external magnetic field ($B$), represents a landmark achievement in topological physics and offers a promising route for low-power-consumption electronics [1,2]. The Chern number $C$ not only characterizes the bulk topology of QAH insulators but also dictates the number of dissipationless chiral edge channels [3]. The pursuit of topological materials hosting high and gate-tunable $C$ is thus crucial for advancing topological electronics, as it would enable multi-channel quantum transport, the investigation of topological phase transitions, and the exploration of exotic fractional topological states [4-8]. While the QAH effect has been realized in several materials, such as magnetically doped topological insulators [2], intrinsic magnetic topological insulators [9,10], and moiré systems [11-16], most of them exhibit a Chern number of $|C| = 1$. Tunable-$C$ QAH can be achieved by stacking decoupled $C = 1$ QAH multilayers separated by normal insulator spacers [17]. However, experimental access to designed QAH states with intrinsic high-$C$ topological bands remains elusive.

Rhombohedral multilayer graphene has emerged as a promising platform for engineering various QAH states, owing to its layer-dependent band topology [18-20]. Furthermore, its intrinsic layer-dependent surface flat bands, which follow an approximate energy dispersion relation of $E \sim k^N$ (where $N$ is the layer number), host rich correlated topological phases due to the interplay between strong electron correlations and topological band structures [21-27]. Indeed, various topological states, including integer and fractional QAH effects, have been reported in rhombohedral graphene systems [15,28-30]. Much attention has focused on moiré superlattices formed by aligning rhombohedral graphene with hexagonal boron nitride (h-BN) [15,30-33]. Although theoretical studies predict that the valley Chern number of rhombohedral graphene scales with layer number [19,34-36], experiments in rhombohedral graphene/hBN moiré systems have predominantly reported Chern bands with $C = 1$ across different layer numbers [15,30-32,37,38]. In contrast, recent advances in twisted rhombohedral graphene homostructures suggest a more versatile platform for exploring high-$C$ QAH states [39-41]. However, how QAH topology systematically evolves with layer number and layer configuration remains unclear.

In this work, we report the experimental realization of a layer-engineered and electrically reconfigurable platform for high-$C$ QAH insulators in twisted rhombohedral graphene homostructures, featuring layer-dependent topology, gate-tunable chirality, and electric-field-driven topological phase transitions. Through systematic studies of twisted monolayer-rhombohedral $N$-layer (denoted as $(1 + N)$ L) graphene and twisted Bernal bilayer-rhombohedral tetralayer (denoted as $(2 + 4)$ L) graphene, we demonstrate three key advances. First, we observe a systematic layer-dependent evolution of topology in twisted $(1 + N)$ L family ($N = 3, 4, 5$), with robust QAH states exhibiting $C = N$. Second, beyond structural tunability, we demonstrate that the chirality of the QAH state (that is, the sign of $C$) can be switched by both carrier density ($n$) and displacement field ($D$). Third, in twisted $(2 + 4)$ L graphene, we realize a displacement-field-driven topological phase transition between two nontrivial QAH states with $C = 3$ and $C = 4$, highlighting the versatility of this platform for engineering tunable topological states.

**Layer-dependent Chern bands**

Our devices are fabricated using a van der Waals assembly technique, where a monolayer (or bilayer)

graphene is placed on top of a rhombohedral multilayer graphene with a controlled small twist angle $\theta$, achieved via a standard cut-and-stack process [42,43]. The stack is fully encapsulated by top and bottom h-BN dielectrics, along with dual-graphite gates for independent electrostatic control. Throughout assembly, graphene is intentionally misaligned with the h-BN to avoid introducing additional moiré structures. The presence of rhombohedral stacking domains in the final stacks is confirmed by Raman mapping (see Methods and Extended Data Fig. 1). Figure 1a illustrates the device architecture of the twisted $(M + N)$ L graphene family. This dual-gate configuration allows for independent tuning of the $n$ and $D$. The formation of a moiré superlattice in our homostructures is evidenced by the observation of resistivity peaks at full band filling ($\nu = 4$), from which $\theta$ is determined and cross-verified via Brown-Zak oscillations (see Methods and Fig. S1).

The band structures for the twisted $(1 + N)$ L graphene, obtained from self-consistent Hartree-Fock calculations based on the continuum model, confirm that the first conduction band is an isolated topological flat band. This band carries a nonzero valley Chern number $C$ that matches the layer number $N$, where $N = 3, 4, 5$ (see Fig. 1b). An optimized interlayer potential $\Delta$ can be experimentally achieved by tuning $D$. Strong Coulomb interactions within this isolated flat band are expected to lift the degeneracy of spin and valley flavors, leading to a spin- and valley-polarized orbital Chern band at odd integer moiré fillings [44,45]. Experimentally, by utilizing the dual-gate structure as illustrated in Fig.1a, we tune the Fermi level to these odd $\nu$ to access the corresponding topologically nontrivial states with $C = N$.

Figure 1c presents representative magnetic hysteresis loops measured at one electron per moiré unit cell ($\nu = 1$) in three devices: twisted monolayer-trilayer ($(1 + 3)$ L, device D1, $\theta = 1.29°$), monolayer-tetralayer ($(1 + 4)$ L, device D2, $\theta = 1.16°$), and monolayer-pentalayer ($(1 + 5)$ L, device D3, $\theta = 1.39°$) graphene. All three devices exhibit a typical QAH effect, characterized by a quantized Hall resistivity $\rho_{xy} = h/Ce^2$ and a vanishing longitudinal resistivity $\rho_{xx}$ down to zero magnetic field. The measured quantized plateaus of $\rho_{xy}$ in device D1, D2, and D3 are $h/3e^2$, $h/4e^2$, and $h/5e^2$, respectively. The observation of QAH states with $C = N$ directly confirms the layer-dependent Chern number predicted by our band calculations and stems from the intrinsic, layer-dependent Berry curvature of rhombohedral graphene [46].

**QAH at $\nu = 1$ in twisted $(1 + 3)$ L and $(1 + 4)$ L graphene**

The full $\nu - D$ map of $\rho_{xx}$ at $B = 0$ T for all devices are provided in Extended Data Fig. 2. In all cases, QAH states emerge when electrons are polarized away from moiré interface by applying a sufficiently large negative $D$. This behavior aligns with prior reports and is analogous to observations in rhombohedral graphene/h-BN systems [15,39].

Detailed measurements near $\nu = 1$ for twisted $(1 + 3)$ L (device D1) and $(1 + 4)$ L (device D2) graphene are shown in Fig. 2a-d and 2e-h, respectively. Figure 2a (2e) present fine $\nu - D$ maps of the symmetrized $\rho_{xx}$ and anti-symmetrized $\rho_{xy}$, measured under a small magnetic field of $B = \pm 0.1$ T ($\pm 90$ mT), in order to stabilize the spontaneous magnetization. In Fig. 2a (device D1), clear minima in $\rho_{xx}$ are observed within the $D$ range of $-0.540 < D < -0.420$ V nm$^{-1}$, accompanied by maxima in $\rho_{xy}$. Similar features appear in Fig. 2e (device D2) for $-0.700 < D < -0.470$ V nm$^{-1}$. These signatures are characteristic of topologically nontrivial states. The higher $|D|$ required to observe the nontrivial states in $(1+4)$ L graphene device, compared to the $(1+3)$ L graphene device, may be attributed to the enhanced interlayer screening effect in thicker graphene, as demonstrated in previous studies [26].

The presence of QAH states at optimized position is further verified by measuring $\rho_{xx}$ and $\rho_{xy}$ as a function of $\nu$ at fixed $D$ (Fig. 2c, g) and as a function of $D$ at fixed $\nu$ (Fig. 2d, h). Well-defined plateau in $\rho_{xy}$ are observed together with vanishing $\rho_{xx}$. The quantized values are $h/3e^2$ for the (1+3) L device and $h/4e^2$ for the (1+4) L device, consistent with the results in Fig. 1c. The Chern number can also be alternatively extracted via the Streda formula, $\frac{\partial n}{\partial B} = C\frac{e}{h}$, by analyzing the Landau fan diagrams from magnetic field-dependent transport. As shown in Fig. 2b and 2f, the slope of these dispersive fan diagrams yields $C = 3$ for the $(1+3)$ L device and $C = 4$ for $(1+4)$ L device, further confirming the corresponding topological state in each case. The QAH effects in our devices are robust, with the decent quantization persisting up to 3 K and Cuire temperature reaching approximately 8 K (see Extended Data Fig. 3).

Beyond the conventional QAH effect at the integer filling $\nu = 1$, we observe that anomalous $\rho_{xy}$ extends into neighboring non-integer filling ranges: for the $(1+3)$ L device in the region $1 < \nu < 1.6$, and for the $(1+4)$ L device in $0.6 < \nu < 1$ and $1 < \nu < 1.8$, as shown in Fig. 2a,e. Notably, at $\nu = 3/2$, the signature of a Chern insulator with $C = 2$ is identified (see Extended Data Fig. 5), indicating an interaction-induced doubling of the moiré unit cell [47]. This unconventional Chern insulator suggests that simultaneous breaking of both time-reversal and discrete translational symmetries, likely originating from topological charge order [16,47].

**Switchable chirality at $\nu = 3$ in twisted $(1+3)$ L graphene**

In addition to the QAH effect at $\nu = 1$, we observe QAH states near $\nu = 3$ with $|C| = 3$ in the twisted $(1+3)$ L device (device D1), as summarized in Fig. 3. Intriguingly, the chirality, namely the sign of $C$, can be effectively switched by either electrostatic doping (effectively the moiré filling factor $\nu$) or the displacement field $D$.

Figure 3a-d display the $\nu - D$ map of $\rho_{xx}$ and $\rho_{xy}$ near $\nu = 3$ with $\nu$ swept forward and backward, measured under a small magnetic field of $B = 0.1$ T. A QAH state with $C = -3$ is identified at $\nu = 2.905$ and $D = -0.317$ V nm$^{-1}$, as confirmed by the magnetic hysteresis loop showing a quantized Hall plateaus at $-h/3e^2$ (Fig. 3g) and by Landau fan diagrams consistent with $C = -3$ via the Streda formula (Fig. 3e, f). Strikingly, similar QAH state but with opposite chirality, namely $C = 3$, can be also found at $\nu = 2.947$ and $D = -0.310$ V nm$^{-1}$ (Fig. 3h). Notably, the precise quantization of $\rho_{xy}$ in both states persists down to zero magnetic field.

The $\nu - D$ maps measured at fixed small $B$ in Fig. 3b and 3d reveal the coexistence of both large positive and negative $\rho_{xy}$, each corresponding to a minimum in $\rho_{xx}$ (Fig. 3a, c). The boundary between these regions is sharp and depends on both $\nu$ and $D$. Comparing Fig. 3b with 3d, it is obvious that the direction of the $\nu$ scan alters the spatial distribution of positive and negative $\rho_{xy}$ domains. A similar dependence on the $D$ scan direction is also observed (Fig. S2). This sensitivity of the $\rho_{xy}$ sign to scan history manifests directly as prominent hysteresis when plotting $\rho_{xy}$ versus $\nu$ (Fig. 3i) or versus $D$ (Fig. 3j). Importantly, $\rho_{xy}$ in Fig. 3i and 3j remains quantized at either $h/3e^2$ or $-h/3e^2$ before and after these switches. Consequently, in our device D1, the chirality of the QAH state near $\nu = 3$ can be controllably switched by three independent parameters: $B$, $\nu$ and $D$.

In Fig. 3i, the $\nu$-induced switching between opposite quantized $\rho_{xy}$ values occurs at the same $\nu$, forming a nearly rectangular hysteresis loop. Beyond the loop, $\rho_{xy}$ deviate from the quantized values. In contrast, the $D$-induced switching between opposite quantized $\rho_{xy}$ values (Fig. 3j) spans a broader range and can occur over different $D$ values beyond the hysteresis loop. Specifically, along the forward scan direction (solid line), $\rho_{xy}$ first quantizes at $h/3e^2$ for $-0.330 < D < -0.293$ V nm$^{-1}$ (region ① and ②), and then switches to $-h/3e^2$ for $-0.290 < D < -0.270$ V nm$^{-1}$ (region ④). On the reverse scan (dashed line), $\rho_{xy}$ first quantizes at $-h/3e^2$ for $-0.307 < D < -0.270$ V nm$^{-1}$ (region ④ and ③), and then switches to $h/3e^2$ for $-0.334 < D < -0.310$ V nm$^{-1}$ (region ①). Notably, opposite Chern numbers can coexist within a single scan direction of $D$ (region ① and ④), a behavior not observed in previous reports [32].

Electrostatic-doping-induced sign reversal of $\rho_{xy}$ has been reported previously in twisted monolayer-bilayer graphene and related systems [29,32,48]. This behavior is a hallmark of orbital Chern insulators, where the total magnetization comprises contributions from both the bulk orbital magnetization and the chiral edge orbital current [49]. Sweeping the doping level across the Chern insulator gap can cause the edge-state contribution to dominate, reversing the total magnetization. Since the K and K' valleys carry opposite orbital magnetization due to their opposite valley Chern numbers, a sign change in the total magnetization drives a switch in the valley polarization. As a result, the Chern number of orbital Chern insulator, and thus the sign of the quantized $\rho_{xy}$, is reversed. Figure 3k illustrates the free energy stability of valley-polarized states for the four states marked in Fig. 3j. The existence of region ① and ④ indicates that the displacement field can directly modify the energetic stability of valley-polarized states (K and K′ valleys), effectively changing the Chern number by selectively stabilizing different valley index depending on the amplitude of $D$.

**Displacement-field-induced Chern number switching in twisted $(2 + 4)$ L graphene**

We further demonstrate that in twisted rhombohedral graphene, the displacement field $D$ can not only switch the sign but also alter the absolute value of the Chern number $C$. This capability is achieved by engineering the layer configuration of the twisted stack. As shown in Fig. 4, we observe in twisted $(2 + 4)$ L graphene (device D4, $\theta = 1.21^\circ$) that $D$ can drive a topological phase transition between a $C = 3$ QAH state and a $C = 4$ QAH state.

Figure 4b presents the $\nu - D$ map of symmetrized $\rho_{xx}$ and anti-symmetrized $\rho_{xy}$ at $B = \pm 90$ mT. Similar to the twisted $(1 + N)$ L graphene family, topologically nontrivial states emerge near $\nu = 1$, signaled by regions of anomalously large $\rho_{xy}$ and vanishing $\rho_{xx}$. A distinct feature, however, is that the $\rho_{xx}$ minima are split into two separate regions by a local maximum near $D \approx -0.628$ V nm$^{-1}$. Meanwhile, the corresponding $\rho_{xy}$ plateaus quantize to two different values. Figure 4f displays a linecut scan along red vertical line in Fig. 4b. Remarkably, $\rho_{xy}$ is quantized at $h/3e^2$ for $-0.683 < D < -0.641$ V nm$^{-1}$ and at $h/4e^2$ for $-0.608 < D < -0.571$ V nm$^{-1}$.

Figure 4e shows $\rho_{xy}$ and $\rho_{xx}$ as a function of $\nu$ at two characteristic displacement fields $D = -0.669$ V nm$^{-1}$ and $D = -0.602$ V nm$^{-1}$, marked by violet and white horizontal lines in Fig. 4b, respectively. The quantization of $\rho_{xy}$ and vanishing $\rho_{xx}$ are consistent with Fig. 4f. Magnetic hysteresis measurements taken at the positions indicated by green and pink stars in Fig. 4b exhibit pronounced hysteresis loops, with precisely quantized $\rho_{xy} = h/3e^2$ and $\rho_{xy} = h/4e^2$ persisting down to zero magnetic field, respectively, confirming their QAH nature. The Chern numbers are further corroborated by Landau fan diagrams (Fig. 4g, h), whose slopes according to the Streda formula yield $C = 3$ and $C = 4$ at two selected $D$. The temperature evolution of this transition is shown in Extended Data Fig. 4.

Although displacement-field-induced topological phase transitions have been theoretically proposed in various systems [20,44,50,51], their experimental realization in a QAH regime, specifically a transition between two distinct topologically nontrivial Chern states, has remained elusive. Our observation of a $D$-driven topological phase transition from $C = 3$ to $C = 4$ provides this demonstration and agrees well with band structure calculations, as shown in Fig. 4a.

**Discussion and conclusion**

Our observations of layer- and gate-tunable high-$C$ QAH states in twisted rhombohedral graphene establish a versatile platform for both fundamental research in topological matter and future quantum device engineering. The realization of such states opens the door to exploring a rich hierarchy of correlated topological orders beyond the conventional Landau level paradigm. In particular, fractional Chern insulators in high-$C$ Chern bands are promising candidates for hosting non-Abelian excitations, which are of central interest for topological quantum computation [52,53]. Already in our devices, we observe signatures of unconventional Chern insulators emerging at non-integer fillings, including an integer Chern insulator at $\nu = 3/2$ and possible topological signatures near $\nu = 7/3$ in the twisted $(1 + 3)$ L device (see Extended Data Fig. 5, 6). These topologically nontrivial states manifest as an anomalous Hall effect with magnetic hysteresis and dispersive fan diagrams. Although the corresponding $\rho_{xy}$ fails to quantize at zero magnetic field, which we attribute to residual disorder or imperfect contact (see Tab. S1), our system provides a promising platform for approaching the appealing fractional Chern insulators in high-$C$ Chern bands.

The QAH states reported here exhibit well-developed quantization and suppressed longitudinal resistance (see Fig. S5) across seven devices (see the summary of measured devices in Extended Data Fig. 7-9 and Tab. 1), highlighting improved robustness compared to other systems. From the perspective of electronic applications, our work validates a design principle for topological band engineering: the Chern number can be programmed by selecting specific layer configurations, shifting the field from discovery-driven to design-driven research (see our prediction in Extended Data Fig. 10). This principle could be extended to create even higher Chern bands using thicker graphene stacks. Remarkably, the rich choice of QAH states with different $C$, when coupled with proximity-induced superconductivity, offers a route to engineer chiral Majorana edge modes. Furthermore, the ability to electrostatically switch between distinct Chern numbers in a single device introduces a novel mechanism for topological electronics. Such a device could function as a topological transistor, modulating the number of dissipationless edge channels via gate voltage alone.

## Methods

### Device fabrication

Our heterostructures consist of twisted monolayer (or bilayer) graphene and rhombohedral multilayer graphene, encapsulated by top and bottom h-BN dielectric layers, with few-layer graphite serving as the top and bottom gates. Graphite and h-BN flakes were mechanically exfoliated onto $SiO_2$/Si substrates from bulk crystals. The layer numbers of multilayer graphene were determined by optical contrast. Rhombohedral stacking domains were first identified using an infrared camera and then confirmed by Raman spectroscopy (see Extended Data Fig. 1). Monolayer (bilayer) and rhombohedral multilayer graphene regions were selected within the same flake and separated into individual pieces by atomic force microscope (AFM) lithography or ultrafast laser cutting. All stacks were assembled using a standard dry-transfer technique assisted by a poly (bisphenol A carbonate) (PC)/polydimethylsiloxane (PDMS) stamp. The transfer sequence was: topmost h-BN, top graphite, top h-BN, monolayer (bilayer) graphene, and finally the rhombohedral multilayer graphene at a controlled twist angle. The assembled stack was then released onto a prepared bottom stack consisting of bottom graphite and bottom h-BN, which has been cleaned by AFM in contact mode. The presence of rhombohedral stacking domains in the final stack was further verified by additional Raman mapping. AFM imaging was used to identify bubble-free regions for device fabrication. The stacks were then patterned into Hall bar geometries using standard electron-beam lithography followed by reactive-ion etching. Electrical contacts were made by depositing Cr/Au via electron-beam evaporation.

Device D3, which was fabricated with a metallic top gate and a graphite bottom gate, has been systematically studied previously [39]. All other devices in this study are newly fabricated with both top and bottom graphite gates.

### Electrical transport measurements

Measurements were performed in a Helium-4 cryostat with a Helium-3 insert with a base temperature of 0.3 K (Oxford TeslatronPT, for devices D1 and D4) or in a dilution refrigerator with a base temperature of approximately 50 mK (Oxford Triton, for devices D2 and D3), unless otherwise specified. Standard low-frequency lock-in techniques (SR830 or Zurich Instruments MFLI) were employed to measure the longitudinal resistance $R_{xx}$ and Hall resistance $R_{xy}$ at an excitation frequency of 17.7 Hz with an ac current of 1 nA - 5 nA. Gate voltages were applied using Keithley 2614B or Keithley 2450 source-measure units.

The carrier density $n$ and displacement field $D$ in the devices were derived from the top- and bottom- gate voltages ($V_{\mathrm{t}}$ and $V_{\mathrm{b}}$) via: $n = \frac{C_{\mathrm{b}}V_{\mathrm{b}}+C_{\mathrm{t}}V_{\mathrm{t}}}{e} - n_0$ and $D = \frac{C_{\mathrm{b}}V_{\mathrm{b}}-C_{\mathrm{t}}V_{\mathrm{t}}}{2\varepsilon_0} - D_0$, where $C_{\mathrm{b}}$ and $C_{\mathrm{t}}$ are the bottom- and top-gate capacitances calibrated from the quantum oscillations, $e$ is the elementary charge, and $\varepsilon_0$ is the vacuum permittivity. The offsets $n_0$ and $D_0$ account for intrinsic doping and built-in electric field. The moiré filling factor $\nu$ is given by $\nu = 4n/n_s$, where $n_s = 4/A$ is the carrier density at full filling and $A$ is the area of the moiré unit cell.

Throughout this work, positive $D$ is defined as the electric field pointing from the rhombohedral multilayer graphene toward monolayer (bilayer) graphene, as illustrated in Fig. 1a.

The Chern number of the QAH states is determined using two independent approaches: (i) quantized Hall resistance, $\rho_{xy} = h/Ce^2$, and (ii) the Středa formula, $\frac{\partial n}{\partial B} = C\frac{e}{h}$, extracted from Landau fan diagrams based on magnetotransport measurements.

**Twisted angle determination**

The twist angle $\theta$ was extracted from the carrier density $n_s$ at full filling ($\nu = 4$) using the relation $n_s \approx 8\theta^2/\sqrt{3}a^2$, where $a = 0.246$ nm is the graphene lattice constant. Alternatively, $\theta$ was independently determined from Brown-Zak oscillations, where minima in $R_{xx}$ occur at $B = \phi_0/qA$. Here, $\phi_0 = h/e$ is the flux quantum, $q$ is an integer, and $A = \sqrt{3}\lambda^2/2$ is the moiré unit-cell area. Linear fitting of $q$ versus $1/B$ yields the moiré wavelength $\lambda$, from which $\theta$ is calculated. The twist angles of all devices are summarized in Extended Data Tab. 1.

**Symmetrization and anti-symmetrization**

To eliminate geometric mixing between $R_{xx}$ and $R_{xy}$, we applied standard symmetrization and anti-symmetrization procedures. For the data at fixed magnetic field $B$, we employed $R_{xx}(\pm B) = [R_{xx}(+B) + R_{xx}(-B)]/2$ and $R_{xy}(\pm B) = [R_{xy}(+B) - R_{xy}(-B)]/2$. For magnetic hysteresis loops, the following definitions were used: $R_{xy}^{\text{anti-sym}}(B, \leftarrow) = [R_{xy}^{\text{raw}}(B, \leftarrow) - R_{xy}^{\text{raw}}(-B, \rightarrow)]/2$, $R_{xy}^{\text{anti-sym}}(B, \rightarrow) = [R_{xy}^{\text{raw}}(B, \rightarrow) - R_{xy}^{\text{raw}}(-B, \leftarrow)]/2$, $R_{xx}^{\text{sym}}(B, \leftarrow) = [R_{xx}^{\text{raw}}(B, \leftarrow) + R_{xx}^{\text{raw}}(-B, \rightarrow)]/2$, and $R_{xx}^{\text{sym}}(B, \rightarrow) = [R_{xx}^{\text{raw}}(B, \rightarrow) + R_{xx}^{\text{raw}}(-B, \leftarrow)]/2$, where arrows indicate the scan direction of $B$. Resistivity were obtained as $\rho_{xx} = R_{xx}W/L$ and $\rho_{xy} = R_{xy}$, with $W$ and $L$ being the width and length of the Hall bar, respectively.

**Band structure calculation based on continuum model**

The single-particle band structure was calculated by using continuum model. Specifically, the Hamiltonian of twisted rhombohedral multilayer graphene is

$$\mathcal{H}_{tot} = \mathcal{H}_{m+n} + \mathcal{T} + V_{ext},$$

where $\mathcal{H}_{m+n} = h_{m,-\theta/2} + h_{n,\theta/2}$ is the Hamiltonian describing the twisted graphene formed by the bottom ($m$-layer) and top ($n$-layer) stacks, $\mathcal{T}$ encodes the hopping processes between the bottom and top stacks, and $V_{ext}$ describes the external displacement-field-induced potential on the each layer of twisted graphene. $h_{N,\phi}$ denotes the Hamiltonian of rotated rhombohedral $N$-layer graphene, with the rotation angle $\phi$ relative to the $x$-axis and constructed from Slonczewski-Weiss-McClure tight-binding lattice model. Mathematical details and physical parameters can be found in Supplementary Information.

**Self-consistent Hartree-Fock calculation**

The interacting band structure was calculated by further applying self-consistent Hartree-Fock method to the single-particle band structure. The Coulomb interaction Hamiltonian is given by

$$H^{\text{int}} = \frac{1}{2A}\sum_{\alpha,\alpha'}\sum_{\tilde{k}_1,\tilde{k}_2,\tilde{q}} V_{\alpha\alpha'}(\tilde{q}) c^{\dagger}_{\alpha,\tilde{k}_1+\tilde{q}} c^{\dagger}_{\alpha',\tilde{k}_2-\tilde{q}} c_{\alpha',\tilde{k}_2} c_{\alpha,\tilde{k}_1}$$

where $A$ is the total area of the sample, $\alpha = A_1, \dots, B_{N_b+N_t}$ corresponds to the mixture of sublattice and layer indices, $\tilde{k}$, $\tilde{q}$ are defined in the big Brillouin zone, and the dual gate-screened Coulomb potential

$$V_{\alpha\alpha'}(q) = \frac{e^2 \tanh(|q| d_s)}{2\varepsilon_0 \varepsilon_r |q|}$$

with the screening length $d_s = 53$ nm for $(1+5)$ L and 28 nm for other configurations to emulate the experiment and the relative dielectric constant $\varepsilon_r = 20$. The first five conduction bands were used to calculate. The details can be found in Supplementary Information. Notably, in certain configurations and parameter regimes, the single-particle Hamiltonian alone yields an isolated, flat, and topologically nontrivial first conduction band. For completeness, we include Hartree–Fock terms in our calculations and find that the band topology remains unchanged.

*Note added: During the preparation of this manuscript, we became aware of other complementary and independent studies on twisted Bernal bilayer-rhombohedral tetralayer graphene [54] and twisted monolayer-rhombohedral multilayer graphene family [55].*

**Data availability**

The data that support the findings of this study are available from the corresponding author upon reasonable request.

**Acknowledgements**

This work was funded by National Natural Science Foundation of China (Grant No. 12550402, 12274354, 12574203, 12474144, 92577107, and 22473099), the Zhejiang Provincial Natural Science Foundation of China (Grant No. LR24A040003, LR26B010001 and XHD23A2001), Hangzhou Natural Science Foundation for Key Program (Grant No. 2025SZRJJ1093) and Westlake Education Foundation at Westlake University. We thank Chao Zhang from the Instrumentation and Service Center for Physical Sciences (ISCPS) at Westlake University for technical and facility support in data acquisition. We also thank the Instrumentation and Service Center for Molecular Sciences (ISCMS) at Westlake University for facility support. K.W. and T.T. acknowledge support from the JSPS KAKENHI (Grant Nos. 21H05233 and 23H02052) and World Premier International Research Center Initiative (WPI), MEXT, Japan.

**Author contributions**

S.X. conceived the idea and supervised the project. Z.C. and N.L. fabricated the devices with the assistance of H.X., W.Zhou, X.F., L.W., Q.C., and X.C.. Z.C. and N.L. performed transport measurements with the assistance of J.D. and W.Zhou. N.L., Z.C., N.X., W.Zhu, and S.X. performed data analysis. J.H. and W.Zhu calculated the band structure. K.W. and T.T. grew h-BN crystals. N.L., C.Z., J.H., W.Zhu, and S.X. wrote the manuscript. All authors contributed to the discussions.

## Figures

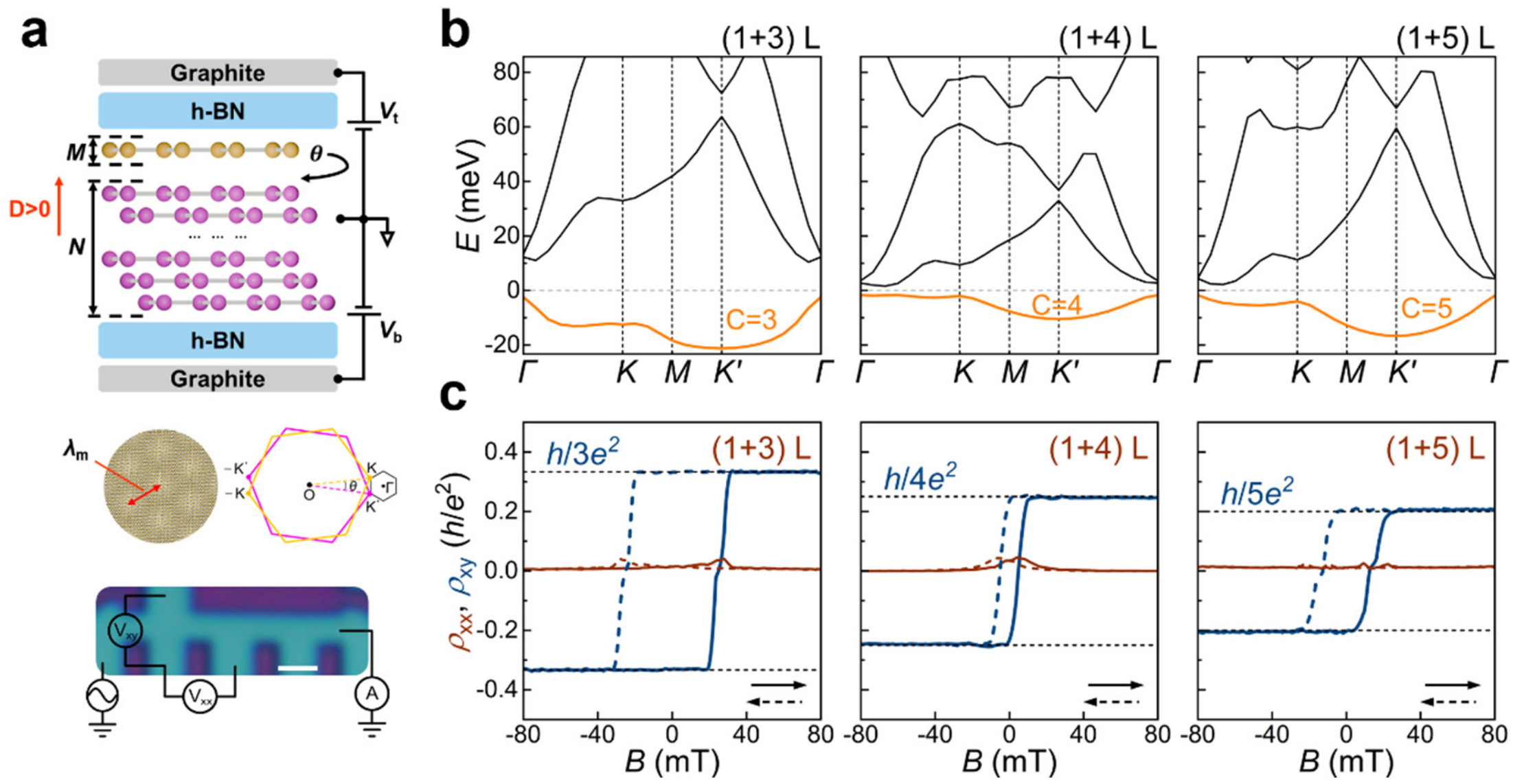

**Fig. 1 | Layer-dependent QAH effect in twisted monolayer-rhombohedral multilayer graphene family. a**, Device schematic with dual graphite gates. A moiré superlattice forms between monolayer (or bilayer) graphene and rhombohedral multilayer graphene with a twist angle $\theta$. The $D$ is defined as positive when pointing from the rhombohedral graphene toward monolayer (bilayer) graphene. Bottom panel shows optical image of a final device with scale bar of 1 μm. **b**, Calculated band structures of twisted $(1 + N)$ L graphene based on self-consistent Hartree-Fock method. Energies are measured with respect to the Fermi level. The twist angles (interlayer potential differences) used in the calculation are $\theta = 1.29°$ ($\Delta = 9.0$ meV), $\theta = 1.16°$ ($\Delta = 12.5$ meV), $\theta = 1.39°$ ($\Delta = 11.0$ meV) for $N = 3, 4, 5$, respectively (see details in Methods). In each case, the first conduction band is an isolated Chern band with the valley Chern number $C$, which coincides with the layer number $N$. **c**, Magnetic hysteresis of symmetrized $\rho_{xx}$ and anti-symmetrized $\rho_{xy}$ as a function of $B$ swept forward and backward at $\nu = 1$ for twisted $(1 + N)$ L graphene devices with $N = 3$ (device D1, $D = -0.470$ V nm$^{-1}$), $N = 4$ (device D2, $D = -0.508$ V nm$^{-1}$), and $N = 5$ (device D3, $D = -0.620$ V nm$^{-1}$). Data were taken at $T = 0.3$ K for device D1 and $T = 50$ mK for device D2 and D3.

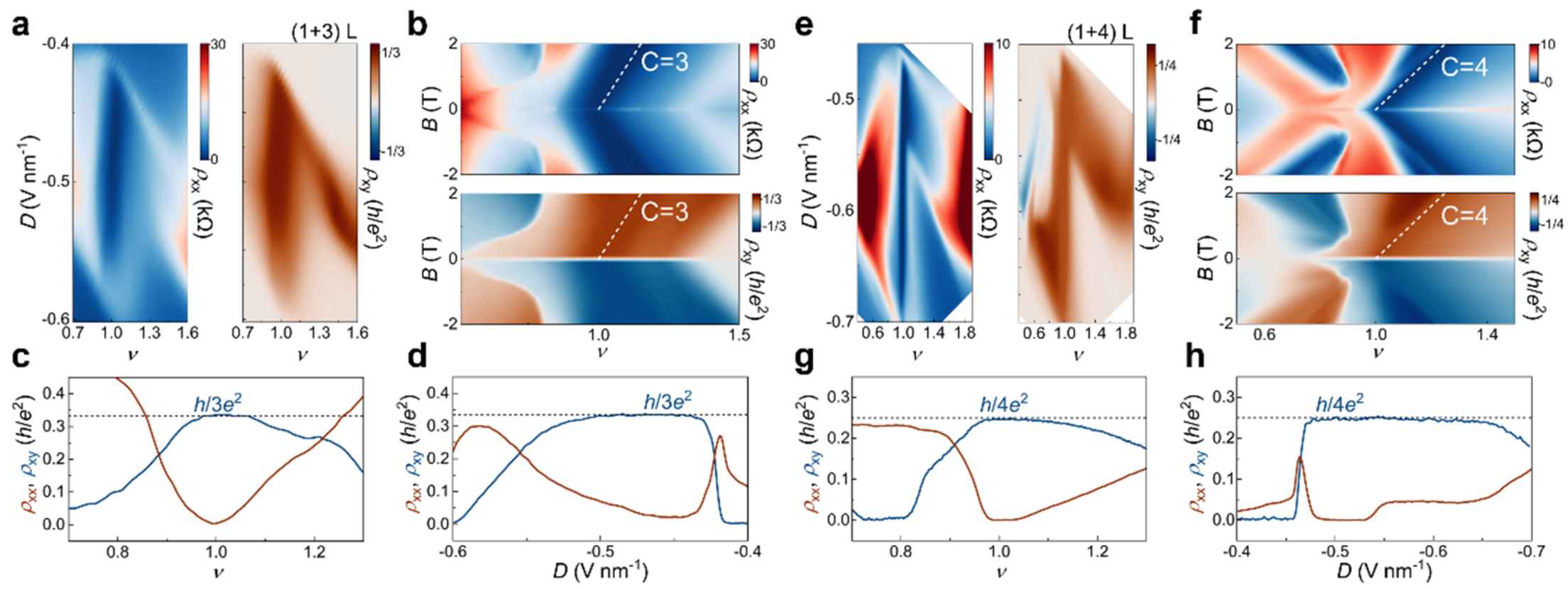

**Fig. 2 | QAH effects with $C = 3$ ($C = 4$) at $\nu = 1$ in twisted $(1 + 3)$ L ($(1 + 4)$ L) graphene. a-d**, Data from twisted $(1 + 3)$ L graphene device (device D1, $\theta = 1.29°$) measured at $T = 0.3$ K. **a**, Symmetrized $\rho_{xx}$ (left) and anti-symmetrized $\rho_{xy}$ (right) as functions of $\nu$ and $D$ measured at $B = \pm 0.1$ T. **b**, Landau fan diagrams of symmetrized $\rho_{xx}$ (top) and anti-symmetrized $\rho_{xy}$ (bottom) at fixed $D = -0.470$ V nm$^{-1}$. White dashed lines trace the dispersions of the minima in $\rho_{xx}$ and plateaus in $\rho_{xy}$. Their slope gives $C = 3$ via the Streda formula. **c**, **d**, Symmetrized $\rho_{xx}$ and anti-symmetrized $\rho_{xy}$ versus $\nu$ at fixed $D = -0.470$ V nm$^{-1}$ **(c)** and versus $D$ at fixed $\nu = 1$ **(d)** measured at $B = \pm 50$ mT. Black dashed lines mark the quantized plateaus $\rho_{xy} = h/3e^2$. **e-h**, Corresponding data from twisted $(1 + 4)$ L graphene device (device D2, $\theta = 1.16°$) measured at $T = 50$ mK. **e**, Symmetrized $\rho_{xx}$ (left) and anti-symmetrized $\rho_{xy}$ (right) versus $\nu$ and $D$ measured at $B = \pm 90$ mT. **f**, Landau fans of symmetrized $\rho_{xx}$ (top) and anti-symmetrized $\rho_{xy}$ (bottom) at fixed $D = -0.508$ V nm$^{-1}$. White dashed lines yield $C = 4$. **g**, **h**, Symmetrized $\rho_{xx}$ and anti-symmetrized $\rho_{xy}$ versus $\nu$ at fixed $D = -0.508$ V nm$^{-1}$ (**g**) and versus $D$ at fixed $\nu = 1$ (**h**) measured at $B = \pm 90$ mT. Black dashed lines indicate the quantized plateaus $\rho_{xy} = h/4e^2$.

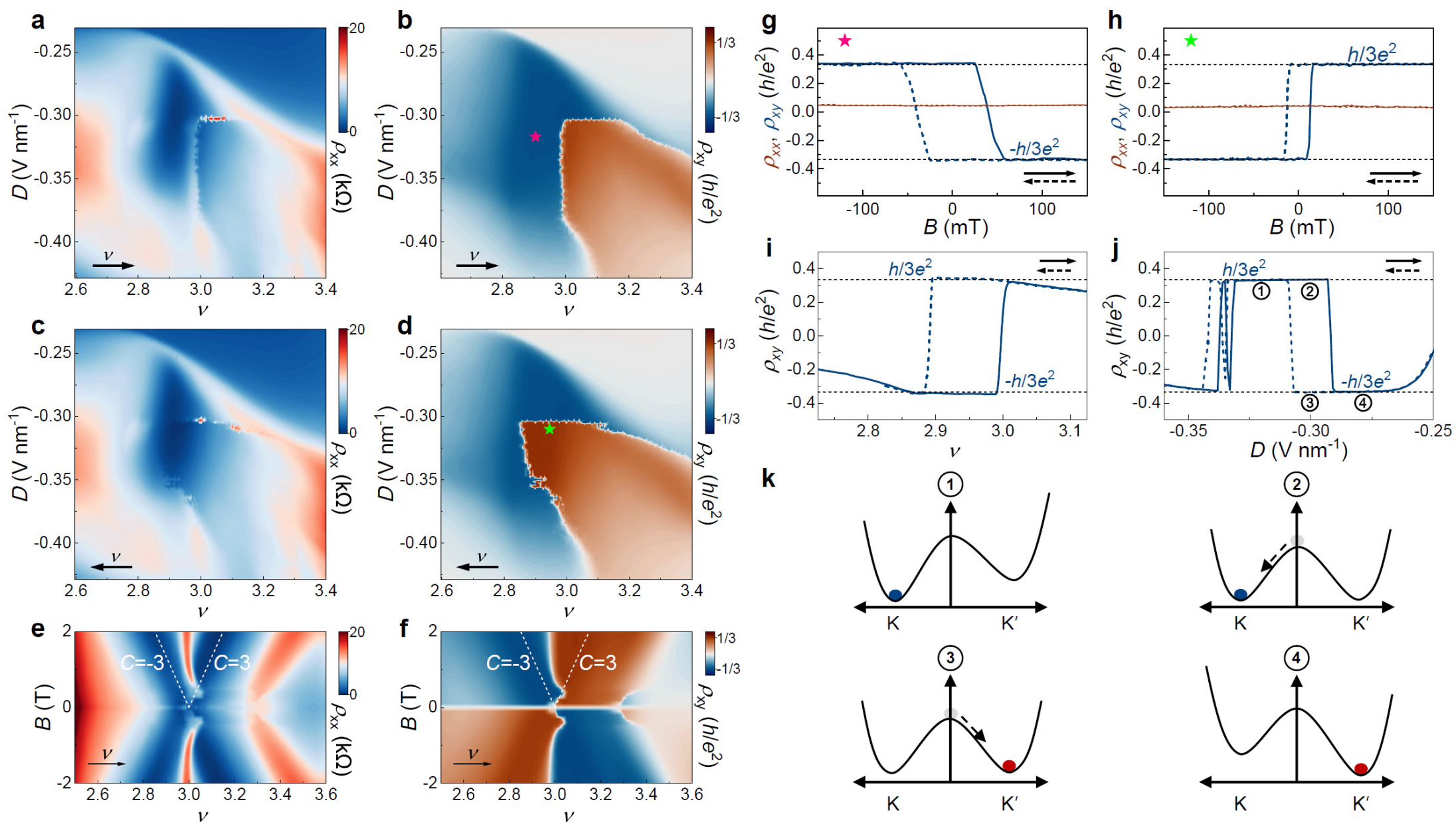

**Fig. 3 | Electrically switchable chirality of QAH states near $\nu = 3$ in twisted $(1+3)$ L graphene. a**, **b**, $\rho_{xx}$ (**a**) and $\rho_{xy}$ (**b**) as functions of $\nu$ and $D$ with $\nu$ swept forward measured at $B = 0.1$ T. **c**, **d**, Corresponding maps with $\nu$ swept backward. **e**, **f**, Landau fan diagrams of symmetrized $\rho_{xx}$ (**e**) and anti-symmetrized $\rho_{xy}$ (**f**) at fixed $D = -0.317$ V nm$^{-1}$. White dashed lines trace the dispersions of $\rho_{xx}$ minima and $\rho_{xy}$ plateaus, yielding two Chern states of $C = 3$ and $C = -3$ via the Streda formula. **g**, **h**, Magnetic hysteresis of symmetrized $\rho_{xx}$ and anti-symmetrized $\rho_{xy}$ as a function of $B$ swept forward and backward measured at the positions marked by the pink star ($\nu = 2.905$ and $D = -0.317$ V nm$^{-1}$, **g**) and green star ($\nu = 2.947$ and $D = -0.310$ V nm$^{-1}$, **h**) in (**b**) and (**d**), respectively. **i**, $\rho_{xy}$ as a function of $\nu$ swept forward and backward at fixed $D = -0.317$ V nm$^{-1}$ and $B = 0.1$ T. **j**, $\rho_{xy}$ as a function of $D$ swept forward and backward at fixed $\nu = 2.947$ and $B = 0.1$ T. Black dashed lines in (**i**) and (**j**) mark the quantized plateaus $\rho_{xy} = \pm h/3e^2$. **k**, Schematic of the free energy versus valley order parameter for four magnetic states marked in (**j**). All data were acquired from device D1 at $T = 0.3$ K.

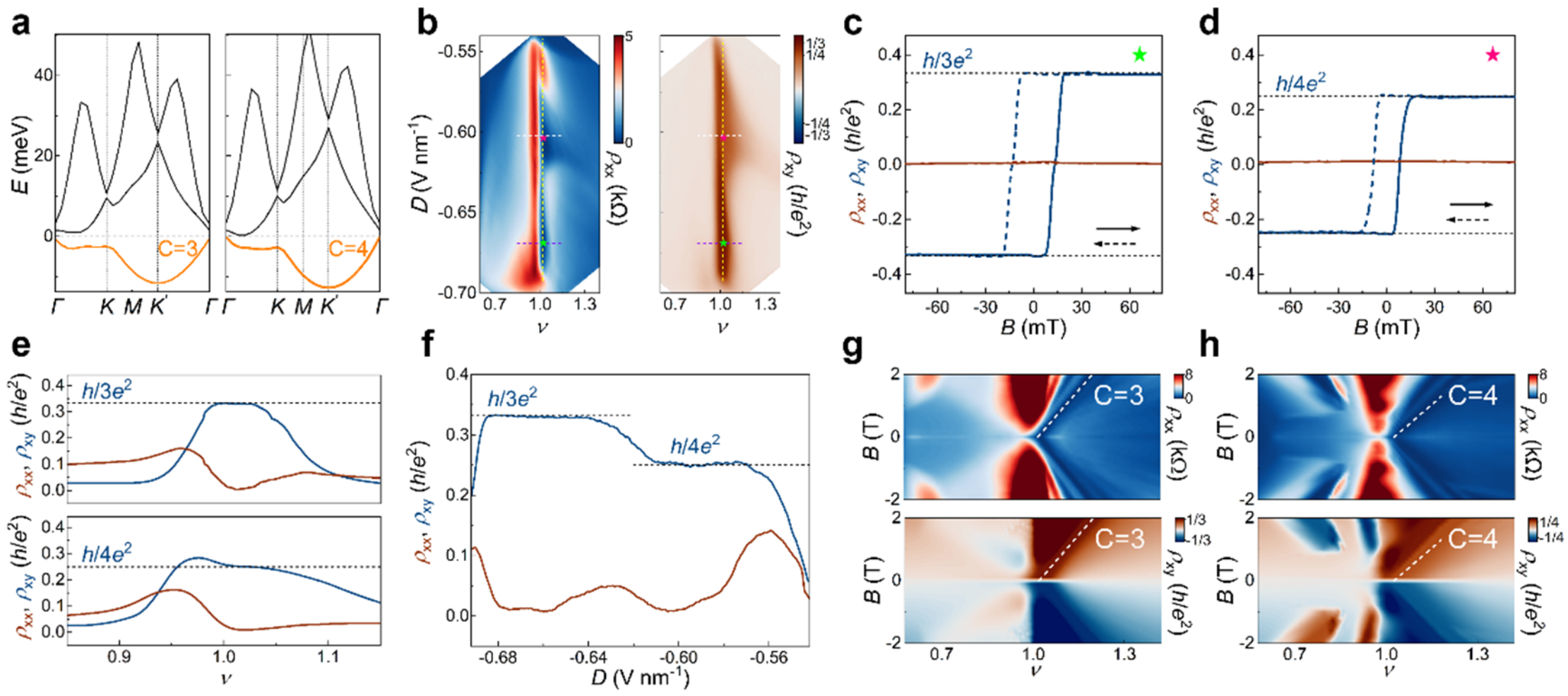

**Fig. 4 | Displacement-field-induced topological phase transition at $\nu = 1$ in twisted $(2+4)$ L graphene.** **a**, Self-consistent Hartree-Fock band structures under interlayer potential differences $\Delta = 11.5$ meV (left) and $\Delta = 10.0$ meV (right). **b**, Symmetrized $\rho_{xx}$ (left) and anti-symmetrized $\rho_{xy}$ (right) versus $\nu$ and $D$ measured at $B = \pm 90$ mT. **c**, **d**, Magnetic hysteresis of symmetrized $\rho_{xx}$ and anti-symmetrized $\rho_{xy}$ as a function of $B$ swept forward and backward measured at the green star ($\nu = 1.02$ and $D = -0.669$ V nm$^{-1}$) (**c**) and pink star ($\nu = 1.02$ and $D = -0.603$ V nm$^{-1}$) (**d**) marked in (**b**). **e**, $\rho_{xx}$ and $\rho_{xy}$ versus $\nu$ at fixed $D = -0.669$ V nm$^{-1}$ (top, violet line in **b**) and $D = -0.602$ V nm$^{-1}$ (bottom, white line in **b**) measured at $B = \pm 90$ mT. **f**, $\rho_{xx}$ and $\rho_{xy}$versus $D$ at fixed $\nu = 1.02$ measured at $B = \pm 90$ mT. Data were measured along yellow vertical line indicated in (**b**). Black dashed lines in (**e**) and (**f**) mark the quantized plateaus corresponding to $h/3e^2$ or $h/4e^2$. **g**, **h**, Landau fan diagrams of $\rho_{xx}$ (top) and $\rho_{xy}$ (bottom) at fixed $D = -0.669$ V nm$^{-1}$ (**g**) and $D = -0.602$ V nm$^{-1}$ (**h**). White dashed lines yield $C = 3$ (**g**) and $C = 4$ (**h**) via the Streda formula. All data were acquired from device D4 at $T = 0.3$ K.

## Extended Data Figures

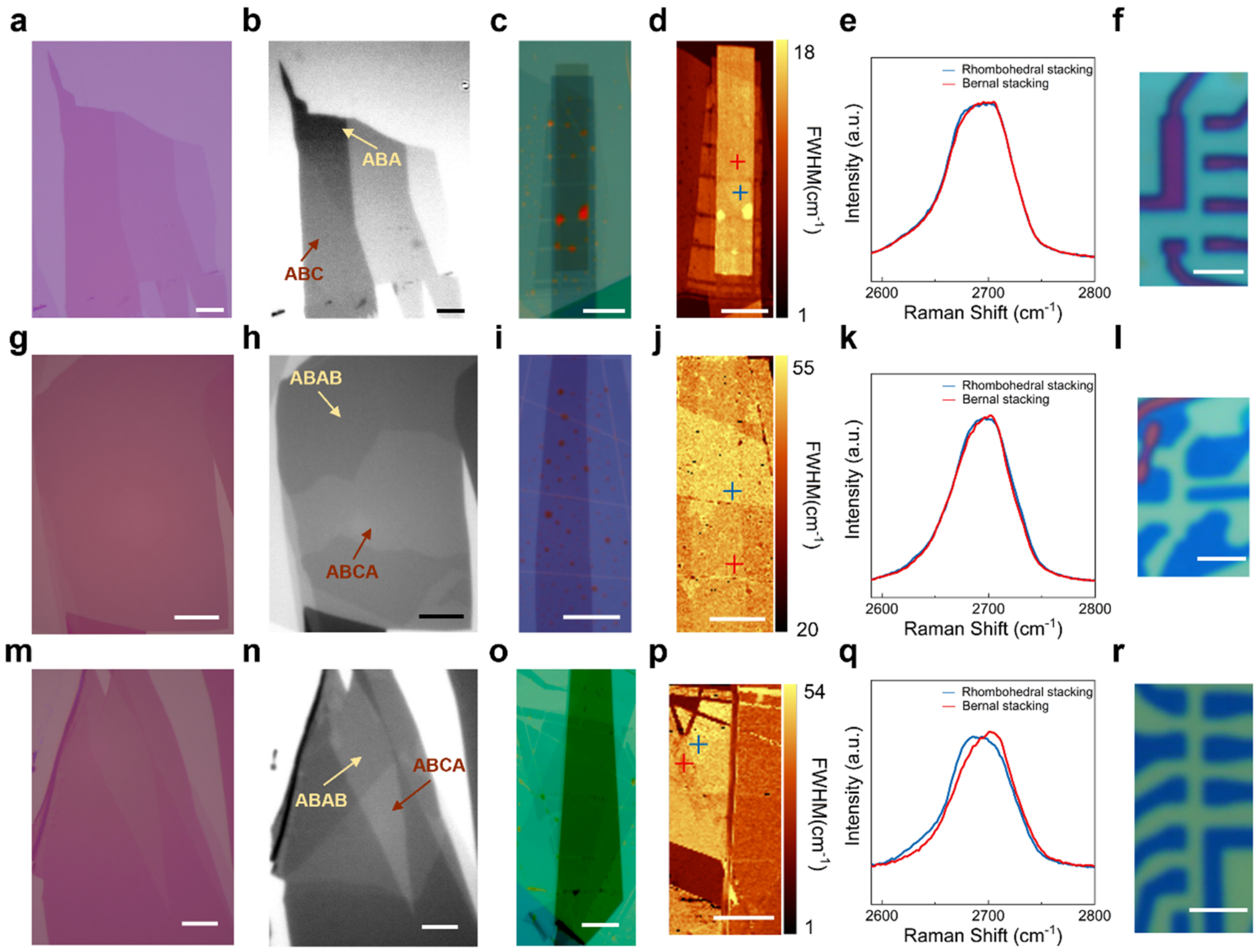

**Extended Data Fig. 1 | Optical Images and Raman characterization of twisted rhombohedral graphene devices. a-f**, Data for twisted (1 + 3) L graphene (device D1). **g-l**, Data for twisted (1 + 4) L graphene (device D2). **m-r**, Data for twisted (2 + 4) L graphene (device D4). **a**, **b**, **g**, **h**, **m**, **n**, Optical and corresponding infrared images of rhombohedral trilayer (**a**, **b**) and tetralayer (**g**, **h**, **m**, **n**) graphene before laser cutting. The Bernal- and rhombohedral-stacking domains in panels (**b**), (**h**), and (**n**) are indicated by yellow and red arrows, respectively. **c**, **d**, **i**, **j**, **o**, **p**, Optical images (**c**, **i** ,**o**) and Raman maps (**d**, **j**, **p**) of the final stacks. **e**, **k**, **q**, Raman spectra of the 2D peak taken at the red and blue crosses in (**d**), (**j**), and (**p**). **f**, **l**, **r**, Optical images of completed devices. Scale bars: 10 μm in (**a-d**, **g-j**, **m-p**); 2 μm (**f**, **l**, **r**).

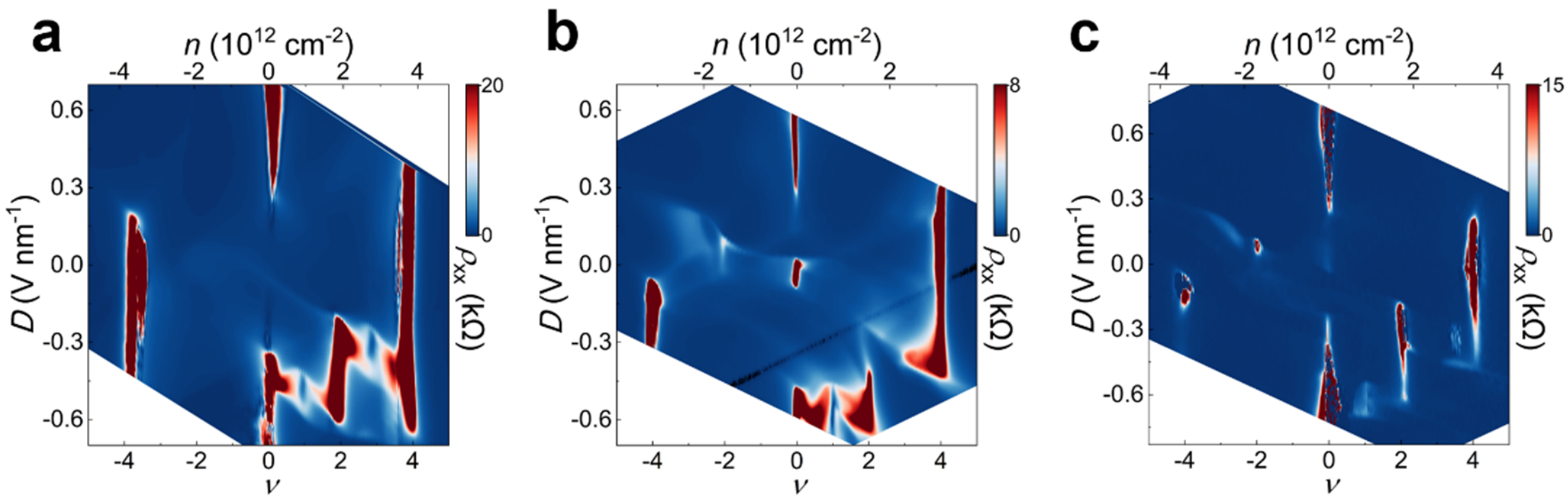

**Extended Data Fig. 2 | Full $\nu - D$ maps at zero magnetic field. a-c**, $\rho_{xx}$ versus $\nu$ and $D$ measured at $T = 0.3$ K for twisted (1 + 3) L graphene (device D1, **a**), twisted (1 + 4) L graphene (device D2, **b**), twisted (2 + 4) L graphene (device D4, **c**).

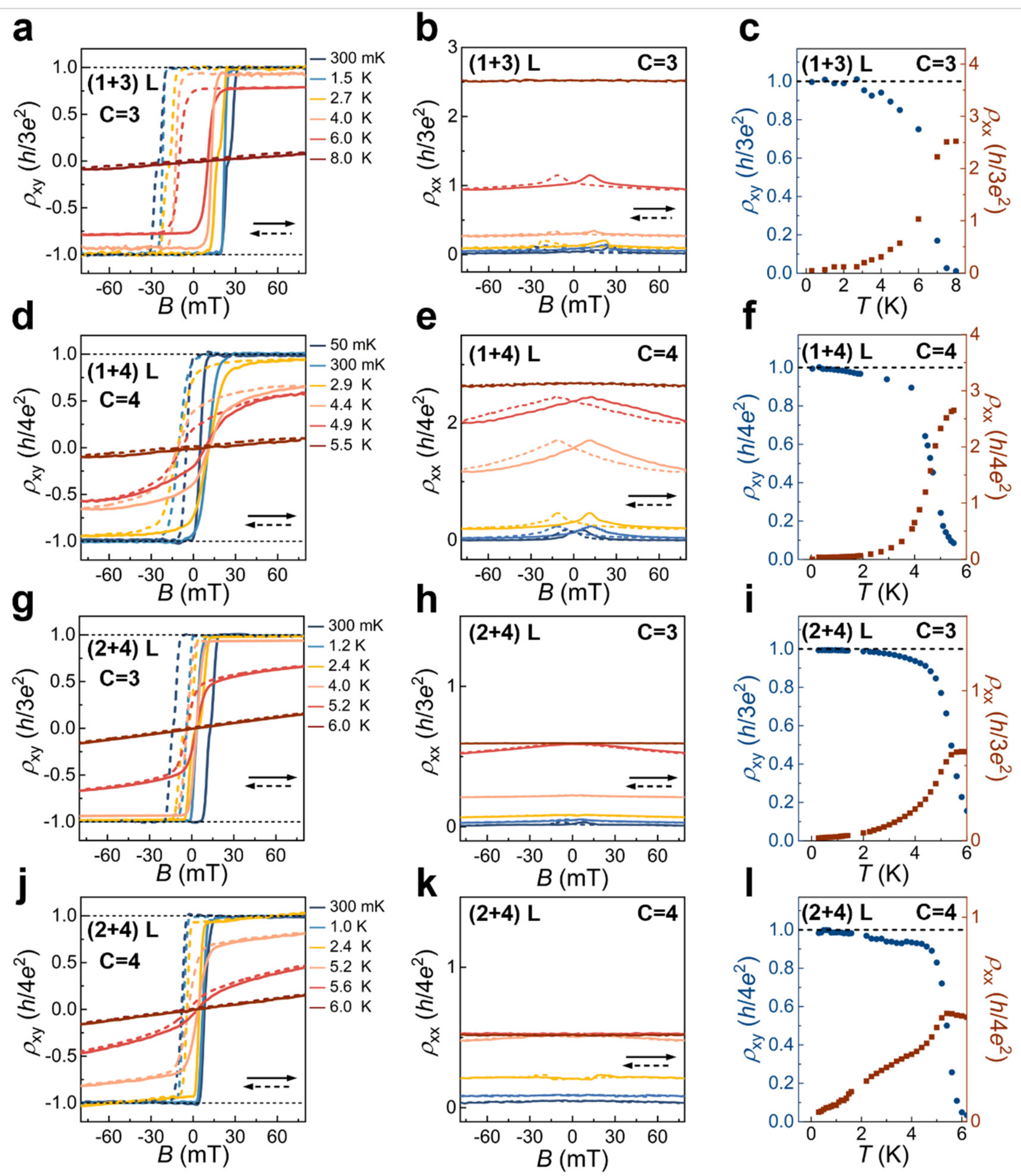

**Extended Data Fig. 3 | Temperature dependence of the QAH states. a-c**, Data for the $C = 3$ state at $\nu = 1$ in device D1. **d-f**, Data for $C = 4$ state at $\nu = 1$ in device D2. **g-i**, Data for $C = 3$ state at $\nu = 1$ in device D4. **j-l**, Data for $C = 4$ state at $\nu = 1$ in device D4. **a**, **b**, **d**, **e**, **g**, **h**, **j**, **k**, Magnetic hysteresis of $\rho_{xx}$ (**a**, **d**, **g**, **j**) and $\rho_{xy}$ (**b**, **e**, **h**, **k**) at selected temperatures. **c**, **f**, **i**, **l**, Extracted $\rho_{xx}$ and $\rho_{xy}$ as a function of temperature.

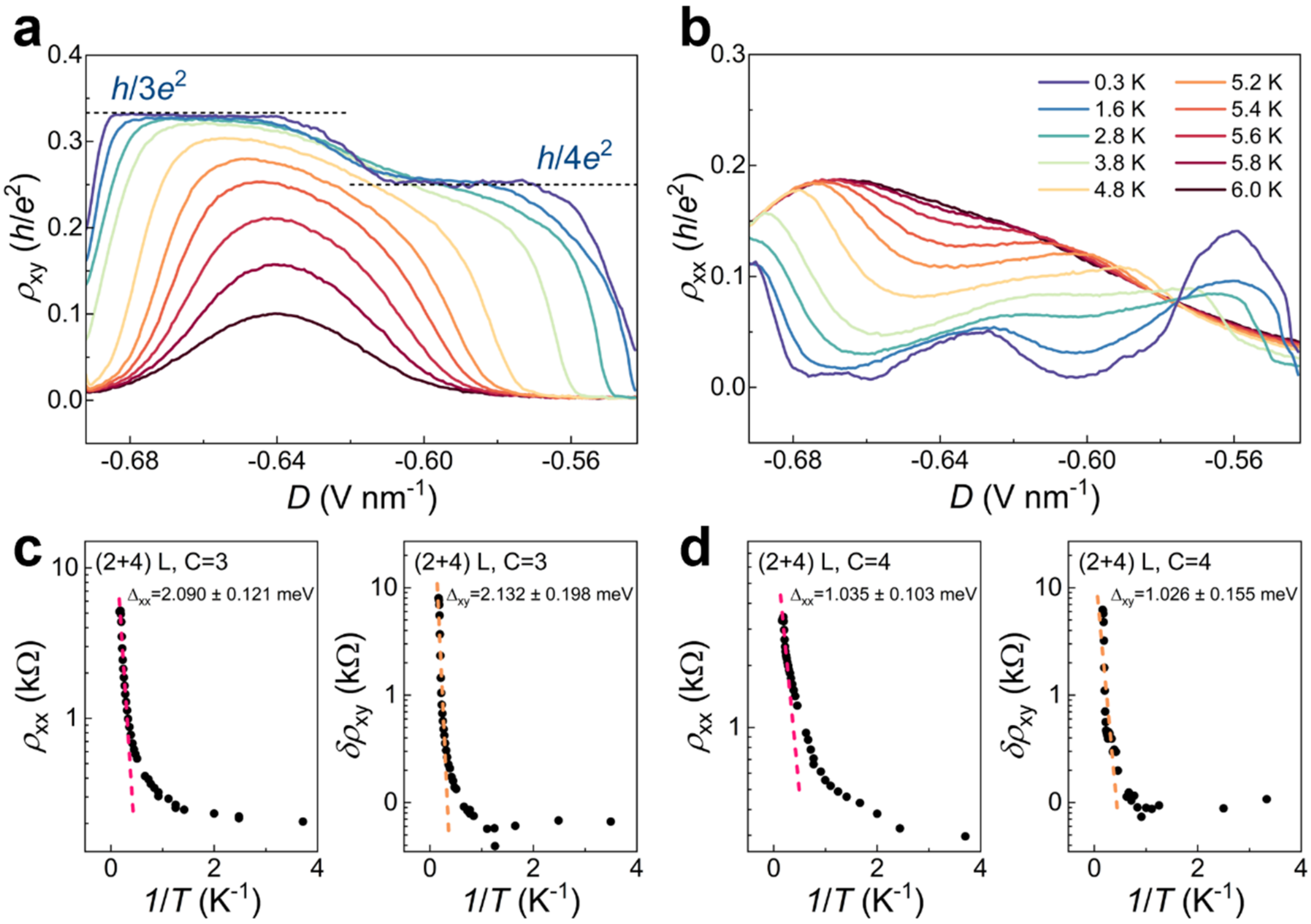

**Extended Data Fig. 4 | Temperature dependence of topological phase transition in device D4.** **a**, **b**, symmetrized $\rho_{xx}$ (**a**) and anti-symmetrized $\rho_{xy}$ (**b**) versus D at fixed $\nu = 1.02$ measured at selected temperatures ($B = \pm 90$ mT). **c**, **d**, Arrhenius plots of $C = 3$ and $C = 4$ states. Extracted symmetrized $\rho_{xx}$ (left panel) and anti-symmetrized $\rho_{xy}$ (right panel) versus $1/T$ for $C = 3$ (**c**) and $C = 4$ (**d**) QAH states at $B = 30$ mT. Thermal activation gaps ($\Delta_{xx}$ and $\Delta_{xx}$) can be extracted from temperature dependence of $\rho_{xx}$ and $\delta\rho_{xy} = h/(Ce^2) - |\rho_{xy}|$. Red (yellow) dashed lines denote linear fit to $\rho_{xx0}e^{-\Delta_{xx}/2k_BT}$ ($\rho_{xy0}e^{-\Delta_{xy}/k_BT}$), where $\rho_{xx0}$ and $\rho_{xy0}$ are fitting constants and $k_B$ is the Boltzmann constant. The results yield the gaps of $\Delta_{C=3} \approx 2.1$ meV and $\Delta_{C=4} \approx 1.0$ meV.

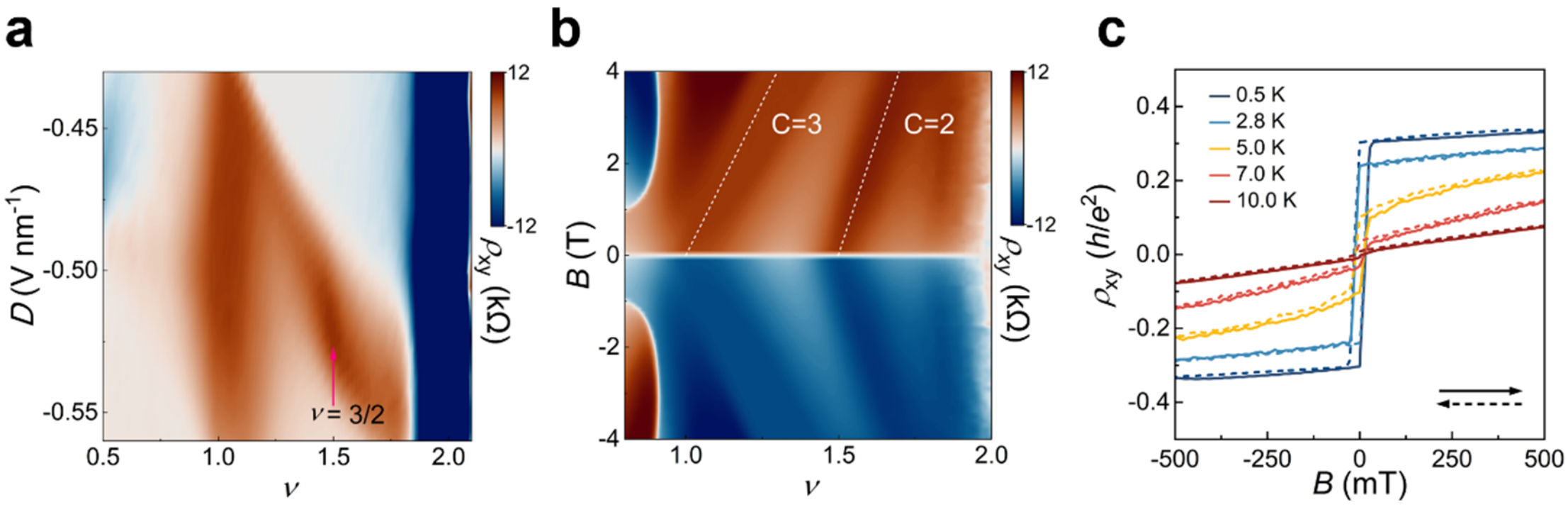

**Extended Data Fig. 5 | Unconventional Chern insulator at $\nu = 3/2$ in twisted $(1 + 3)$ L graphene (device D1).** **a**, $\rho_{xy}$ versus $\nu$ and $D$ measured at $B = \pm 0.1$ T. **b**, Landau fan diagram of $\rho_{xy}$ versus $\nu$ and $B$ at $D = -0.520$ V nm$^{-1}$. **c**, Magnetic hysteresis of $\rho_{xy}$ versus $B$ at $\nu = 3/2$ and $D = -0.520$ V nm$^{-1}$ at selected temperatures.

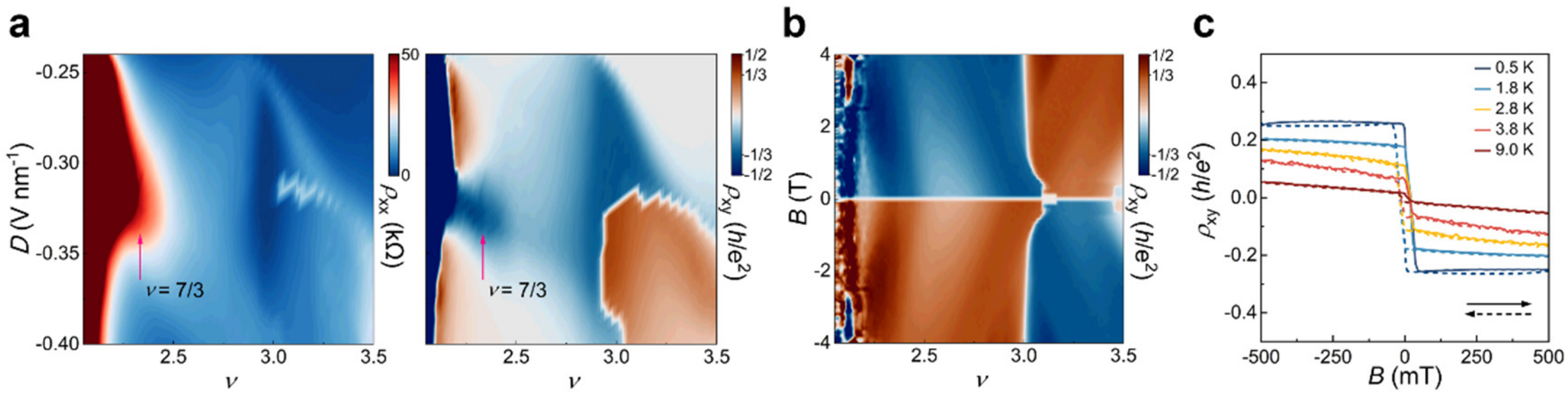

**Extended Data Fig. 6 | Possible topological states near $\nu = 7/3$ in twisted $(1 + 3)$ L graphene (device D1).** **a**, $\rho_{xx}$ (left) and $\rho_{xy}$ (right) as a function of $\nu$ and $D$ measured at $B = \pm 0.1$ T. **b**, Landau fan diagram of $\rho_{xy}$ versus $\nu$ and $B$ at $D = -0.330$ V nm$^{-1}$. **c**, Magnetic hysteresis of $\rho_{xy}$ versus $B$ near $\nu = 7/3$ and $D = -0.330$ V nm$^{-1}$ at selected temperatures.

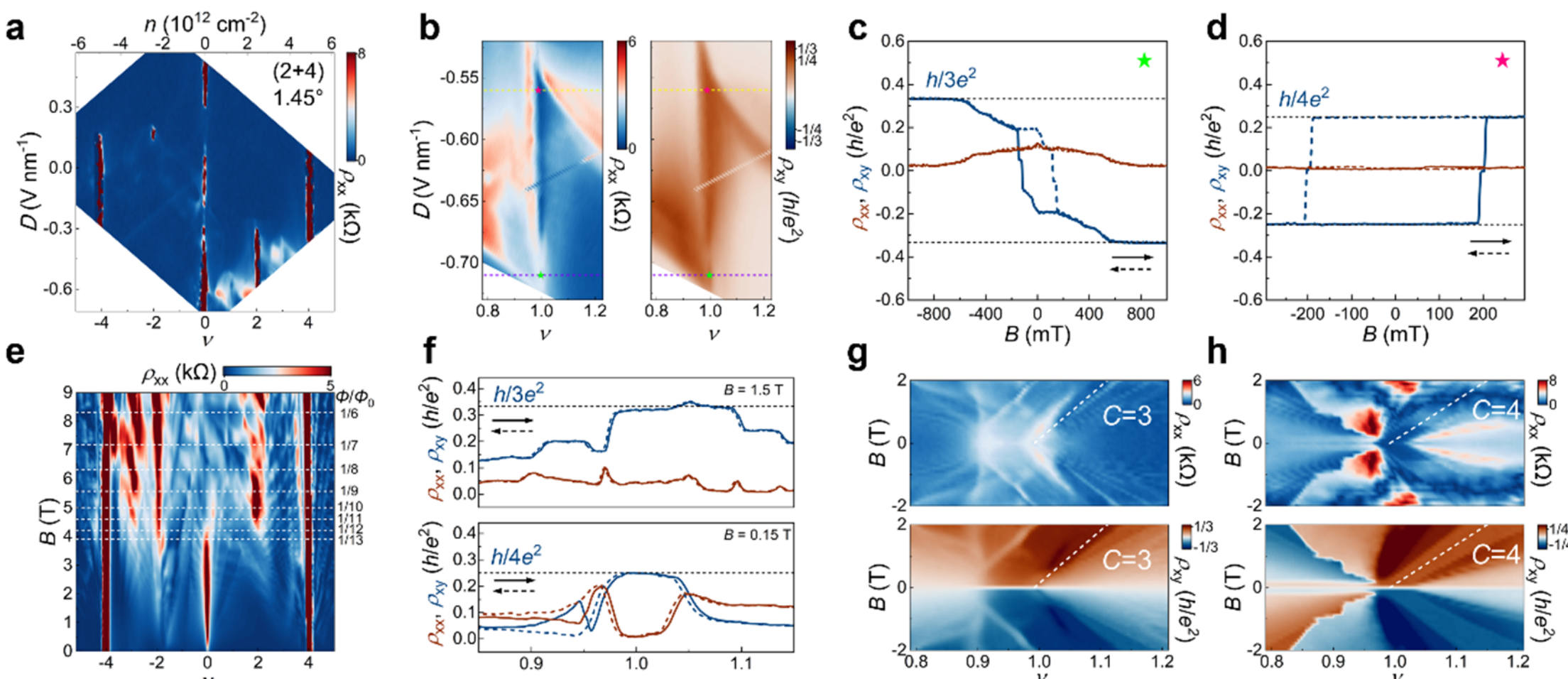

**Extended Data Fig. 7 | Topological states in a twisted (2+4) L graphene with $\theta = 1.45°$ (device D5).** **a,** Full $\nu - D$ map at zero magnetic field. **b**, Symmetrized $\rho_{xx}$ (left) and anti-symmetrized $\rho_{xy}$ (right) as functions of $\nu$ and $D$ measured at $B = \pm 0.1$ T. **c**, **d**, Magnetic hysteresis of symmetrized $\rho_{xx}$ and anti-symmetrized $\rho_{xy}$ as a function of $B$ swept forward and backward, measured at the green star ($\nu = 1.0$ and $D = -0.710$ V nm$^{-1}$) (**c**) and pink star ($\nu = 1.0$ and $D = -0.560$ V nm$^{-1}$) (**d**) marked in (**b**). **e**, Landau fans of $\rho_{xx}$ versus $\nu$ and $B$ at $D = 0$ V nm$^{-1}$. Horizontal white dashed lines mark minima in Brown-Zak oscillations. The extracted moiré wavelength is about 9.69 nm, corresponding to $\theta = 1.45°$. **f**, Symmetrized $\rho_{xx}$ and anti-symmetrized $\rho_{xy}$ versus $\nu$ at fixed $D = -0.560$ V nm$^{-1}$ (bottom, $B = 0.15$ T) and $D = -0.710$ V nm$^{-1}$ (top, $B = 1.5$ T), marked by yellow and violet lines in (**b**), respectively. **g**, **h**, Landau fan diagrams of symmetrized $\rho_{xx}$ (top) and anti-symmetrized $\rho_{xy}$ (bottom) at fixed $D = -0.710$ V nm$^{-1}$ (**g**) and $D = -0.560$ V nm$^{-1}$ (**h**). White dashed lines yield $C = 3$ (**g**) and $C = 4$ (**h**) via the Streda formula. All data were acquired at $T = 0.3$ K.

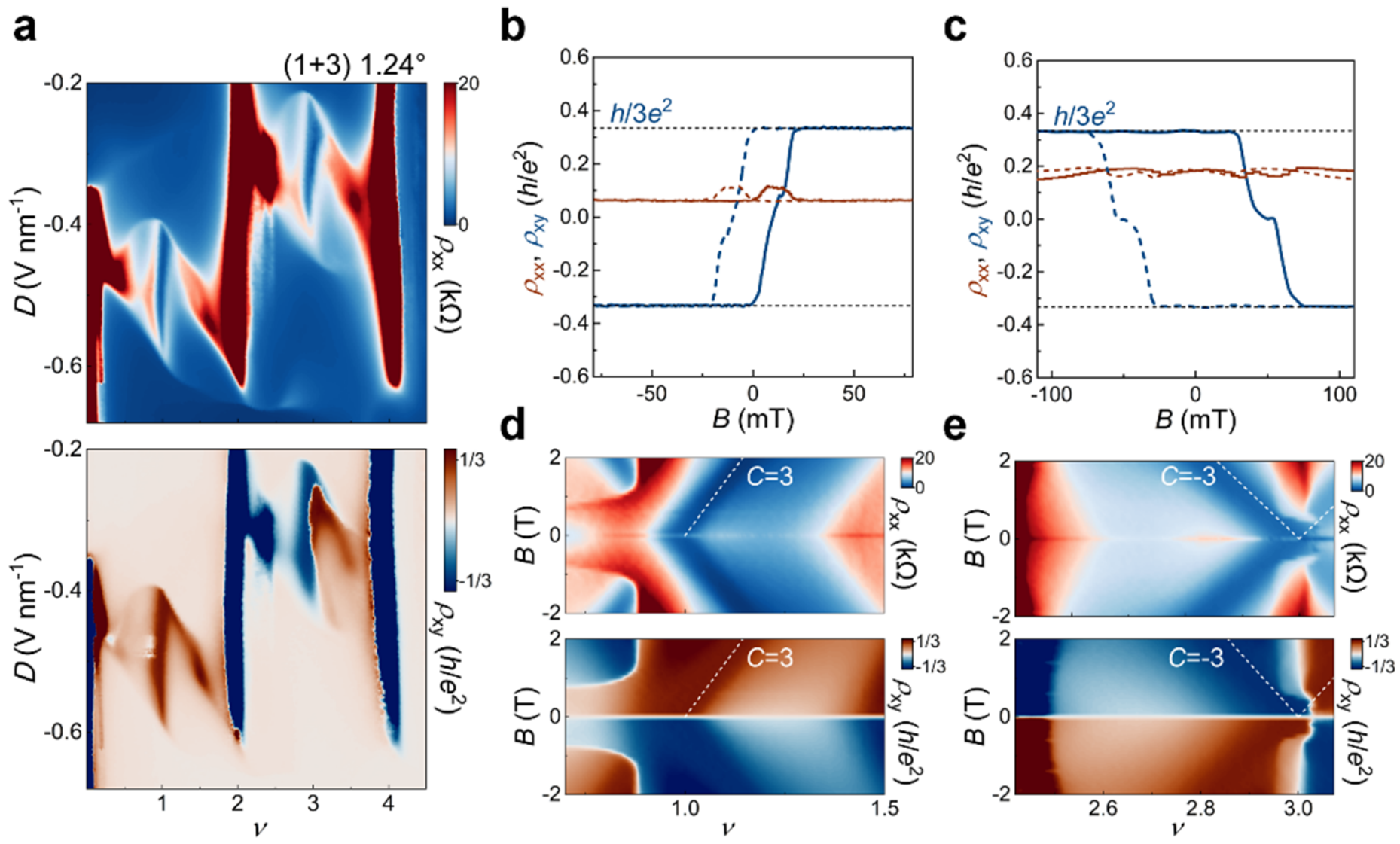

**Extended Data Fig. 8 | Topological states in a twisted (1+3) L graphene with $\theta$ = 1.24° (device D6). a**, Symmetrized $\rho_{xx}$ (top) and anti-symmetrized $\rho_{xy}$ (bottom) as functions of $\nu$ and $D$ measured at $B = \pm 0.1$ T. **b**. **c**, Magnetic hysteresis of symmetrized $\rho_{xx}$ and anti-symmetrized $\rho_{xy}$ as a function of $B$ swept forward and backward measured at $\nu = 1$, $D = -0.440$ V nm$^{-1}$) (**b**) and $\nu = 3$, $D = -0.295$ V nm$^{-1}$ (**c**). **d**, **e**, Landau fan diagrams of symmetrized $\rho_{xx}$ (top) and anti-symmetrized $\rho_{xy}$ (bottom) at fixed $D = -0.510$ V nm$^{-1}$ (**d**) and $D = -0.293$ V nm$^{-1}$ (**e**). White dashed lines yield $C = 3$ (**d**) and $C = -3$ (**e**) via the Streda formula. All data were acquired from device D6 at $T = 0.3$ K.

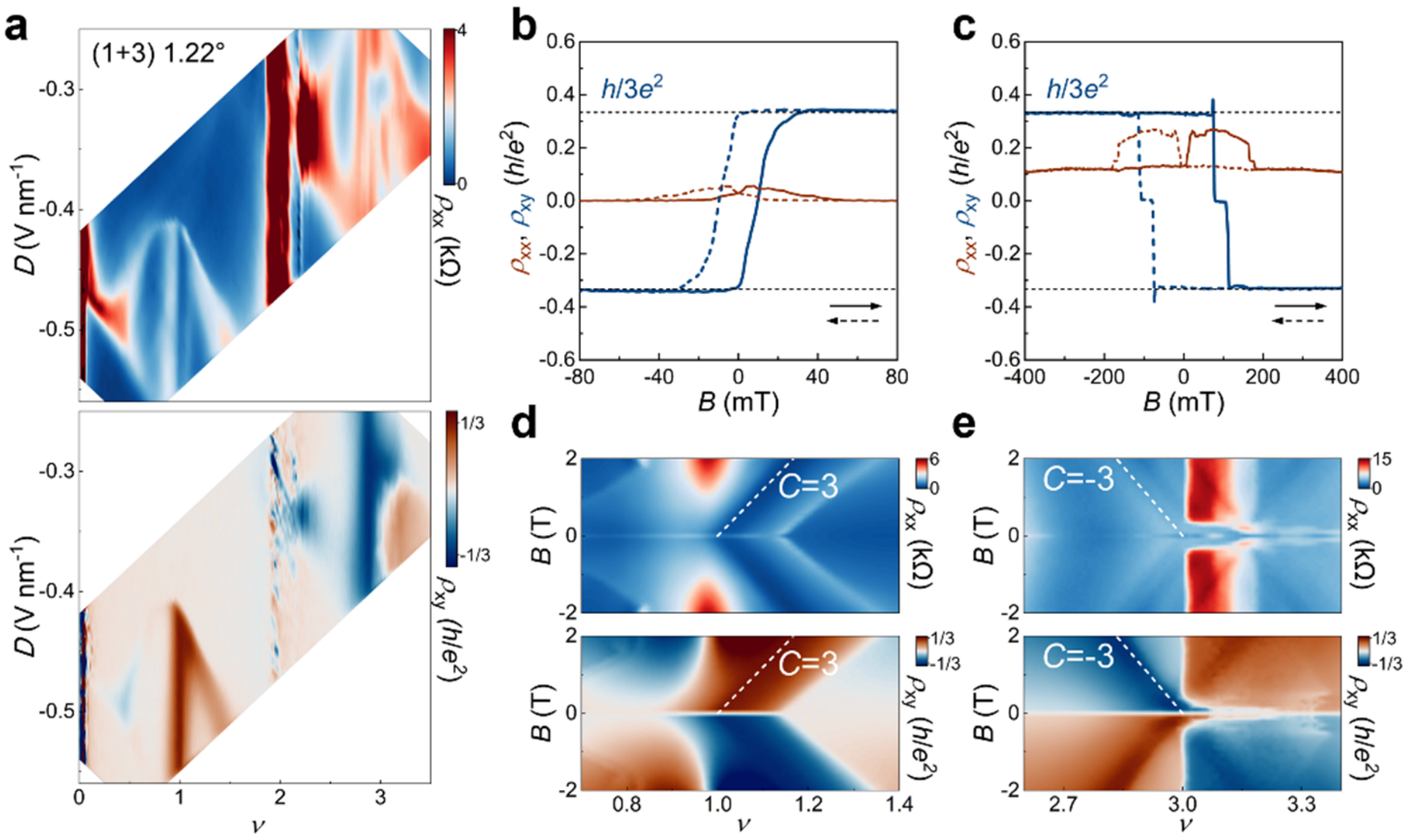

**Extended Data Fig. 9 | Topological states in a twisted (1+3) L graphene with $\theta$ = 1.22° (device D7). a**, Symmetrized $\rho_{xx}$ (top) and anti-symmetrized $\rho_{xy}$ (bottom) as functions of $\nu$ and $D$ measured at $B = \pm 0.1$ T. **b**. **c**, Magnetic hysteresis of symmetrized $\rho_{xx}$ and anti-symmetrized $\rho_{xy}$ as a function of $B$ swept forward and backward measured at $\nu = 1$, $D = -0.441$ V nm$^{-1}$) (**b**) and $\nu = 2.88$, $D = -0.\ 324$ V nm$^{-1}$ (**c**). **d**, **e**, Landau fan diagrams of symmetrized $\rho_{xx}$ (top) and anti-symmetrized $\rho_{xy}$ (bottom) at fixed $D = -0.435$ V nm$^{-1}$ (**d**) and $D = -0.330$ V nm$^{-1}$ (**e**). White dashed lines yield $C = 3$ (**d**) and $C = -3$ (**e**) via the Streda formula. Data in (**b**)(**c**) were acquired at $T = 0.3$ K. Others were acquired at $T = 1.6$ K.

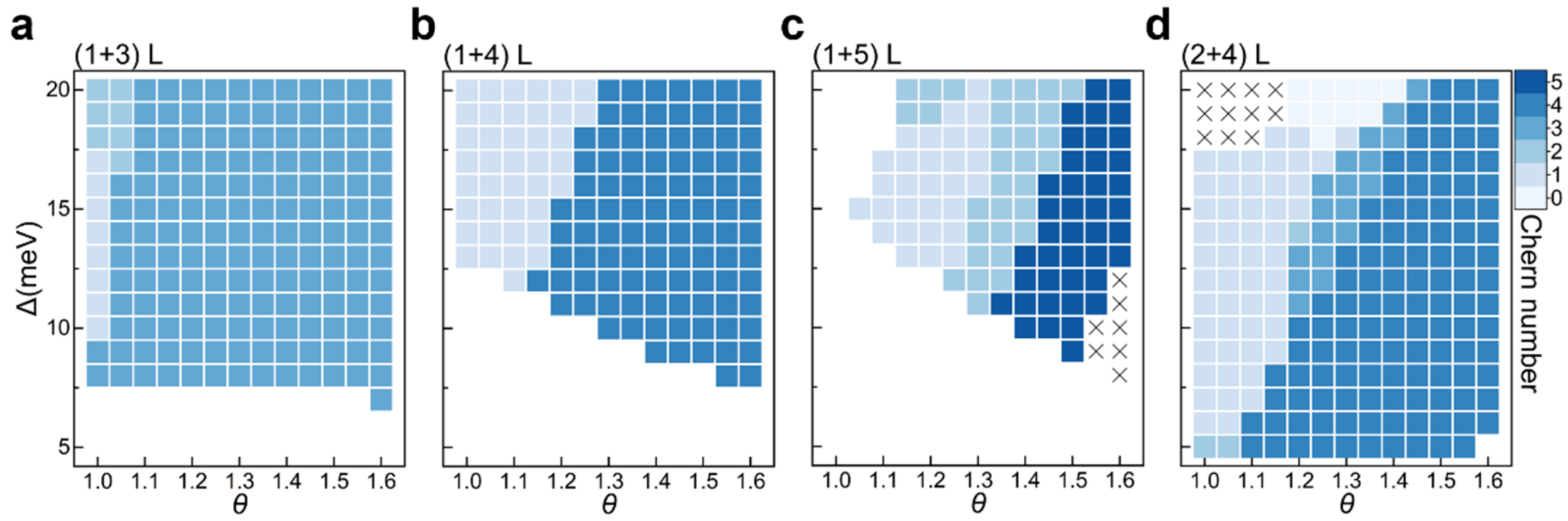

**Extended Data Fig. 10 | Calculated valley Chern numbers evolved with twist angle and interlayer potential difference for various layer configurations. a-d**, The Chern numbers of the first conduction band as functions of twist angle $\theta$ and interlayer potential difference $\Delta$ for twisted $(1+3)$ L (**a**), $(1+4)$ L (**b**), $(1+5)$ L (**c**), and $(2+4)$ L (**d**) graphene. All results are obtained from self-consistent Hartree-Fock calculations based on the continuum model. Data marked by '×' represent a gapless state under given parameters.

**Extended Data Tab. 1| Summary of measured devices.**

| Device | Layer ($M$+$N$) | Twist angle determined by full filling (°) | Moiré wavelength determined by full filling (nm) | Twist angle determined by Brown-Zak oscillations (°) | Moiré wavelength determined by Brown-Zak oscillations (nm) | Observed Chern number |
|---|---|---|---|---|---|---|
| D1 | 1+3 | 1.29 | 10.92 | 1.29 | 10.92 | 3, -3 |
| D2 | 1+4 | 1.15 | 12.26 | 1.16 | 12.23 | 4 |
| D3 | 1+5 | 1.40 | 10.07 | 1.39 | 10.14 | 5 |
| D4 | 2+4 | 1.21 | 11.64 | 1.21 | 11.64 | 3, 4 |
| D5 | 2+4 | 1.45 | 9.69 | 1.45 | 9.69 | 3, 4 |
| D6 | 1+3 | 1.24 | 11.38 | 1.24 | 11.38 | 3, -3 |
| D7 | 1+3 | 1.22 | 11.55 | 1.22 | 11.55 | 3, -3 |